\documentstyle[prl,aps,multicol,epsf]{revtex}
\tighten
\begin{document}
\draft
\title{Suppression of Kondo effect in a quantum dot by
external irradiation
}
\author{A. Kaminski$^1$, Yu.V. Nazarov$^2$, L.I. Glazman$^1$}
\address{$^1$Theoretical Physics Institute, University
of Minnesota, Minneapolis, MN 55455, USA\\
$^2$Delft University of Technology, 2600 GA Delft, The
Netherlands}
\maketitle
\begin{abstract}
We demonstrate that the external irradiation brings
decoherence in the spin states of the quantum dot. This
effect cuts off the Kondo anomaly in conductance even at zero
temperature. We evaluate the dependence of the DC conductance in
the Kondo regime on the power of irradiation, this
dependence being determined by the decoherence.
\end{abstract}
\pacs{PACS numbers: 73.23.Hk, 85.30.Vw, 72.15.Qm}

\begin{multicols}{2}

The Kondo effect have drawn recently a considerable attention in 
connection with the experiments on quantum 
dots\cite{GoldhaberEtal98,CronenwettEtal98,Weis}. Due to the Kondo 
effect, the temperature dependence of the linear conductance across a 
dot becomes non-monotonous:  Upon lowering the temperature, the 
conductance first drops due to the conventional Coulomb blockade, but 
below certain temperature starts growing 
again\cite{GoldhaberEtal98,CronenwettEtal98}. The increase of the 
conductance is associated with the many-body resonance formed at the 
Fermi energy. This resonance manifests itself as a peak in the 
differential conductance $G(V)$ at $V=0$ (zero-bias anomaly) 
\cite{GoldhaberEtal98,CronenwettEtal98,Weis,%
HershfieldEtal91,MeirEtal93}. In a magnetic field, the resonant peak 
in the density of states and therefore the zero-bias peak in $G(V)$ 
are split in two; the inter-peak spacing is proportional to the Zeeman 
energy of the localized 
spin\cite{GoldhaberEtal98,CronenwettEtal98,Weis,MeirEtal93}. These 
results are similar to the effects considered previously in the 
context of tunneling through junctions carrying Kondo 
impurities\cite{Burstein}.

Quantum dot devices are highly controllable, and can be operated in 
regimes inaccessible in the conventional magnetic impurity systems, 
that were used  previously for studying the Kondo effect. Kondo 
anomaly is a  manifestation of a quantum-coherent many-body state. 
Irradiation of a quantum dot with an AC field offers a new, clever way 
of affecting its  dynamics, which enables one to study the Kondo 
anomaly in essentially non-equilibrium conditions. The anomaly 
modified by the irradiation can be investigated by the measurements of 
the DC $I-V$ characteristics.

Despite a considerable amount of 
work\cite{HettlerSchoeller95,Ng96,NordlanderEtal98,LopezEtal98,%
GoldinAvishai97}, the  physical picture of the influence of 
irradiation on the Kondo conductance still needs clarification. 
Nordlander {\it et al.}\cite{NordlanderEtal98} have conjectured that 
the result of irradiation is qualitatively different in two frequency 
domains of the AC field: At sufficiently high frequency, irradiation 
may cause ionization of the quantum dot; loss of the localized spin 
leads to a suppression of the Kondo anomaly in $G(V)$. At frequencies 
below the ionization threshold, irradiation induces satellite 
peaks\cite{HettlerSchoeller95,Ng96,NordlanderEtal98,LopezEtal98} in 
the differential conductance at $eV=n\hbar\omega$, where $\omega$ is 
the frequency of the irradiation. The Kondo effect in these 
conditions, according to \cite{NordlanderEtal98,LopezEtal98}, is not 
suppressed. At zero temperature, it is ``redistributed'' between the 
usual equilibrium Kondo peak at $eV=0$, and its 
satellites\cite{LopezEtal98}: The zero-bias conductance departs from 
the unitary limit, and the satellite peaks appear at its expense; this 
departure is weak as long as the amplitude of AC modulation of the 
dot's energy $eV_{\omega}$ is small.

In this paper we pinpoint the principal effect of the irradiation on  
the Kondo anomaly. This effect consists in the irradiation-induced 
decoherence of the localized spin state.   Contrary to  the picture  
outlined in the previous paragraph, the decoherence occurs even if  
the  irradiation is not able to ionize the dot. We find the dominant  
mechanism of decoherence at the frequencies of AC field below the  
ionization threshold. This mechanism, ``spin-flip cotunneling'', leads  
to a significant deviation of the linear conductance from the  unitary  
limit.  Upon the increase of the AC field frequency to the ionization  
threshold, there is a crossover between the decoherence caused  by 
spin-flip cotunneling and by dot ionization. However, the variation  
of the conductance in this crossover region is parametrically small. 
Starting from fairly low frequencies,  the suppression of the  Kondo 
conductance by decoherence is more important than the  redistribution 
of the conductance  over the high-frequency satellites. 

The system we study is a quantum dot attached to two leads by 
high-resistance junctions so that the charge of  the dot is nearly 
quantized. We describe this system by the Anderson impurity 
Hamiltonian
\begin{mathletters}
\label{Hamiltonian}
\begin{eqnarray}
\hat{H}&=&\!\sum_{k,\sigma,\alpha}(\xi_k+eV_\alpha)
c^{\dagger}_{k\sigma\alpha}c^{\phantom{\dagger}}_{k\sigma\alpha}
+\!\sum_{k,\sigma,\alpha}v_\alpha(
c^{\dagger}_{k\sigma\alpha}d^{\phantom{\dagger}}_{\sigma}+
d^{\dagger}_{\sigma}c^{\phantom{\dagger}}_{k\sigma\alpha}
)\nonumber\\
&+&\sum_\sigma (-E_d+eV_\omega\cos\omega t) d^{\dagger}_{\sigma}
d^{\phantom{\dagger}}_{\sigma}+
Ud^{\dagger}_{\uparrow}d^{\phantom{\dagger}}_{\uparrow}
d^{\dagger}_{\downarrow}d^{\phantom{\dagger}}_{\downarrow}\;;
\label{H}\\
&&\quad V_{L,R}=\pm\textstyle\frac{1}{2}V\;,\quad
\Gamma_\alpha\equiv 2\pi \nu  v_\alpha^2\;.
\label{Gamma}
\end{eqnarray}
\end{mathletters}
Here the first two terms describe non-interacting electrons in the two 
leads ($\alpha=L,R$), and tunneling of free electrons between the dot 
and the leads, respectively; we assume tunneling matrix elements 
$v_\alpha$ are real, without reducing the generality of the 
Hamiltonian. The dot is described by the third and fourth terms of the 
Hamiltonian, $E_d$ and $U-E_d$ are the ionization and the electron 
addition energy, respectively. The tunneling matrix elements 
$v_\alpha$ are related to the widths $\Gamma_\alpha$ by 
Eq.~(\ref{Gamma}), where $\nu $ is the density of states in a lead. 
The external irradiation is applied to the gate, which is coupled to 
the dot capacitively, and modulates the energy of the electron 
localized in the dot. We assume that the leads are DC-biased, 
neglecting the possible ``leakage'' of the irradiating AC field to the 
leads. The generalization onto the case of nonzero AC bias is 
straightforward.

In the present paper we consider the dot in the Kondo regime, 
$U-E_d,\, E_d\gg \Gamma_{L,R}$. We assume the applied DC and AC fields 
are small,  $eV, eV_\omega \ll E_d, U-E_d$. We are primarily 
interested in  the irradiation effects in the domain $\omega < E_d, 
U-E_d$, where neither dot ionization nor the photon-assisted tunneling 
\cite{Kouwenhoven} occur. Under such conditions, one can make the 
Schrieffer-Wolff transformation \cite{SchriefferWolff66} (more 
precisely, its modification for the time-dependent case) to convert 
the Hamiltonian (\ref{H}) to the Kondo form:
\begin{eqnarray}
\hat{H}_K&=&\hat{H}_0+\hat{H}_t\;,\quad
\hat{H}_0=
\sum_{k,\sigma,\alpha}\xi_k^{\phantom{\dagger}}
c^{\dagger}_{k\sigma\alpha}
c^{\phantom{\dagger}}_{k\sigma\alpha}\;,
\label{HK}\\
\hat{H}_t&=&\!\sum_{\parbox{0.3in}{\scriptsize
$k,\sigma,\alpha$\\
$k'\!,\sigma'\!\!,\alpha'$}}\!
{\cal J}_{\alpha\alpha'}(t)
\left(\textstyle\frac{1}{4}\delta_{\sigma\sigma'}+
\hat{S}_j s^j_{\sigma\sigma'}\right)
c^{\dagger}_{k\sigma\alpha}
c^{\phantom{\dagger}}_{k'\sigma'\alpha'}\;,
\nonumber
\end{eqnarray}
where ${\hat{\bf s}}$ and ${\hat{\bf S}}$ are the spin operators of 
the delocalized electrons in the leads and of the electron on the 
isolated level, respectively; we assume summation over the repeating 
indices $j=x,y,z$. The applied bias is accounted for by the time 
dependence of the coupling term $\hat{H}_t$. The Hamiltonian 
(\ref{HK}) operates within the band $-E_d<\xi_k<U-E_d$, see 
Ref.~\cite{Haldane}. The coupling constants ${\cal J}$ are given by
\begin{eqnarray}
&&{\cal J}_{\alpha\alpha'}(t)=
\frac{\sqrt{\Gamma_\alpha\Gamma_{\alpha'}}}{4\pi\nu}
\exp\left[\frac{ie}{\hbar}(V_{\alpha}-V_{\alpha'})t\right]
\nonumber\\&&\times\sum_{n_1,n_2}J_{n_1}
\left(\frac{eV_\omega}{\hbar\omega}\right)
J_{n_2}\left(\frac{eV_\omega}{\hbar\omega}\right)
\exp[i(n_1-n_2)\omega t]\nonumber\\
&&\times\left[\frac{1}{E_d+n_1\hbar\omega}
+\frac{1}{U-E_d+n_1\hbar\omega}\right.\nonumber\\
&&\left. \;\;\;\;\;\;
+\frac{1}{E_d+n_2\hbar\omega}
+\frac{1}{U-E_d+n_2\hbar\omega}
\right]\;, \label{J}
\end{eqnarray}
where $J_n(x)$ are the Bessel functions.

To calculate the differential 
DC conductance $G(V)$, we employ the non-equilibrium Keldysh technique 
in the time representation. In this formalism
\begin{equation}
G(V)=\frac{\partial}{\partial V}\langle
{\sf S}(-\infty,0)\hat{I} {\sf S}(0,-\infty)
\rangle_0\;,\label{G}
\end{equation}
where $\hat{I}$ is the current operator,
and ${\sf S}(t_2,t_1)$ is the evolution matrix determined by
$\hat{H}_t$.

In the perturbation expansion of (\ref{G}) in powers of the
coupling constant ${\cal J}_{\alpha\alpha'}$, the logarithmic
divergences appear starting from the terms of the third order in
${\cal J}_{\alpha\alpha'}$. A representative term has the following
structure:
\begin{eqnarray}
&& \frac{e^2}{\pi\hbar}
\frac{\left[{\cal J}_{LR}^{\rm (0)}\right]^2
{\cal J}_{RR}^{\rm (0)}}{\hbar^3}
\int_{-\infty}^0 dt_1 \int^0_{t_1} dt_2
\langle \hat{S}_j(0)  \hat{S}_k(t_1)  \hat{S}_l(t_2) \rangle
\varepsilon^{jkl}\nonumber\\
&&\qquad\times
\left[t_1 \cos (eVt_1/\hbar) +
t_2\cos (eVt_2/\hbar)\right]\nonumber\\
&&\qquad\times\sum_{k_1,k_2,k_3}G_{k_1}(-t_2)
G_{k_2}(t_2-t_1)\bar{G}_{k_3}(t_1)
\;,\label{typterm}\\
&&\mbox{where}\quad{\cal J}_{\alpha\alpha'}^{\rm (0)}\equiv
\frac{\sqrt{\Gamma_\alpha\Gamma_{\alpha'}}}{\pi\nu\tilde{E}_d},
\quad \tilde{E}_d\equiv\frac{(U-E_d)E_d}{U}\;,\nonumber
\end{eqnarray}
$G_k(t)$ and $\bar{G}_k(t)$ are the time-ordered and anti-time-ordered 
Green functions of free electrons in the leads, and 
$\varepsilon^{jkl}$ is the antisymmetric unit tensor. This and other 
terms of the same structure yield the Kondo divergency in the 
conductance. If there is no spin decoherence, the averages $\langle 
\hat{S}_j(t_1)  \hat{S}_k(t_2)  \hat{S}_l(t_3) \rangle$ are 
independent on time and equal $(i/4)\varepsilon_{jkl}$. The AC field 
introduces decoherence in the dynamics of the impurity spin, which 
results in a decay of the correlation function:
\begin{eqnarray}
\langle \hat{S}_j(t_1) \hat{S}_k(t_2) \hat{S}_l(t_3) \rangle=(i/4)
\varepsilon_{jkl}
\exp (-t_{max}/\tau)\;,
\nonumber\\
t_{max}\equiv\max\{|t_1-t_2|, |t_2-t_3|,
|t_1-t_3|\}\;.
\label{SSS}
\end{eqnarray}
After summing over the electron states $k_i$, performing the
integration over $t_2$ in Eq.~(\ref{typterm}), and
adding up all the cubic in ${\cal J}^{\rm (0)}$ terms, we arrive at
\begin{eqnarray}
&&G^{(3)}(V)=12\pi^2 \frac{e^2}{\pi\hbar}
\nu^3\left[{\cal J}_{LR}^{\rm (0)}\right]^2
\left[{\cal J}_{RR}^{\rm (0)}+{\cal J}_{LL}^{\rm
(0)}\right]\nonumber\\
&&\quad
\times\int^0_{-\infty}\!\! dt\,
\frac{(-t)\cos (eVt/\hbar)\exp (-|t|/\tau)}
{\sinh^2(\pi Tt/\hbar)+(T/D_0)^2}\left(\frac{\pi T}{\hbar}\right)^2.
\label{Gt}
\end{eqnarray}
The effective bandwidth here is $D_0\equiv \sqrt{E_d(U-E_d)}$ 
\cite{Haldane}. In the absence of spin decoherence, the integral in 
Eq.~(\ref{Gt}) equals $\ln(D_0/\max\{T,eV\})$, and diverges 
logarithmically at low temperature and bias, signaling the Kondo 
anomaly. The leading effect of the irradiation is in cutting off this 
divergency. The decay of the spin correlation function (\ref{SSS}) 
makes $G^{(3)}(V)$ finite even at $T,V\to 0$. We will show that the 
spin decoherence by external irradiation does not require ionization 
of the impurity level, and therefore exists at arbitrary low 
frequencies of the applied AC field. The suppressing effect of the 
irradiation on the Kondo conductance, $G^{(3)}\propto \left[{\cal 
J}^{\rm (0)}\right]^3 \ln(D_0\tau/\hbar)$, is not analytic in the 
intensity of the AC field, and cannot be obtained by a finite-order 
perturbation theory.

In the absence of the dot ionization, the decoherence rate can be
calculated with the help of the Hamiltonian (\ref{HK})--(\ref{J}). In
the case of weak modulation, $eV_\omega<\hbar\omega$, it is sufficient
to account for the single-photon processes only. The part of Hamiltonian 
(\ref{HK})--(\ref{J}) responsible for such processes corresponds to four
terms labeled by
$n_1=0$,
$n_2=\pm 1$, and
$n_1=\pm 1$, $n_2=0$ in the sum (\ref{J}), and is given by
\begin{equation}
\hat{H}_t^{(1)}=-\frac{eV_\omega\cos\omega t}{2\tilde{E}_d}
\sum_{\parbox{0.3in}{\scriptsize
$k,\sigma,\alpha$\\
$k'\!,\sigma'\!\!,\alpha'$}}
{\cal J}_{\alpha\alpha'}^{\rm (0)}
\hat{S}_j s^j_{\sigma\sigma'}
c^{\dagger}_{k\sigma\alpha}
c^{\phantom{\dagger}}_{k'\sigma'\alpha'}\;.
\label{Ht1}
\end{equation}
In deriving Eq.~(\ref{Ht1}), we expanded the Bessel
functions of Eq.~(\ref{J}) up to the first order in
$eV_\omega/\hbar\omega$.

\narrowtext
\begin{figure}
\epsfxsize=8.5cm
\centerline{\epsfbox{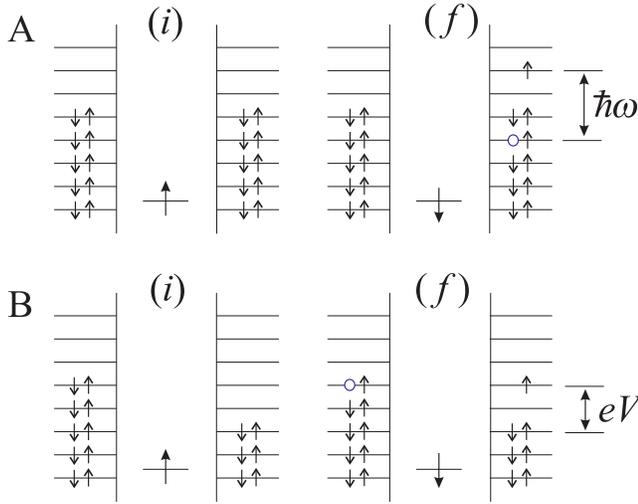}}
\caption{\label{fig:trans}
{\bf Spin-flip cotunneling}: the coherence of the spin state of the 
dot can be lost when an electron interacting with it hops 
from a state below the Fermi level  to a state above the Fermi level. 
Additional energy, needed for such a transition, can be taken either 
from the AC field or from the applied bias. The figure shows the 
initial $(i)$ and final $(f)$ states of the system for the decoherence 
processes driven by an AC field ({\bf A}) and by a finite bias 
({\bf B}).
}
\end{figure}

The process of spin-flip cotunneling (spin flip without ionization of 
the dot) induced by the irradiation is shown schematically in 
Fig.~\ref{fig:trans}A. In terms of the Kondo Hamiltonian (\ref{HK}), 
an electron, which interacts with the dot spin, absorbs a photon and 
hops to a state above the Fermi level, while the spin of the dot 
flips. Within the lowest-order perturbation theory, the rate of this 
process can be calculated with the Fermi Golden Rule applied to 
Hamiltonian (\ref{Ht1}):
\begin{equation}
\frac{\hbar}{\tau}=\frac{1}{2\pi}\hbar\omega
\left[\frac{\Gamma_L+\Gamma_R}{\tilde{E}_d}\right]^2
\left[\frac{eV_\omega}{\tilde{E}_d}\right]^2\; .
\label{Gammad}
\end{equation}
The spin-flip cotunneling persists at arbitrary
low frequencies, leading to the decoherence of the dot spin state.

As we pointed out  earlier, the Kondo anomaly is a
manifestation of a  quantum-coherent many-body state. The loss of spin
coherence suppresses the Kondo anomaly. At $T,V\to 0$, it is the spin
decoherence time $\tau$ what cuts off  the logarithmic  divergency in
the integral (\ref{Gt}). After the first logarithmic correction
(\ref{Gt}) to the conductance is found, we can proceed with the
derivation of the Renormalization Group equation, which yields the
conductance $G$ in the leading logarithm approximation. For the
present non-equilibrium problem, we have to modify the ``poor man's''
technique\cite{Anderson} in order to apply it directly to $G$, rather
than to the scattering amplitudes. This need  emerges from the 
kinetic nature of the problem at hand.   The resulting formula for the
peak conductance, which is valid in the domain  $\hbar/\tau\gtrsim
T_K$, can be cast in the form
\begin{equation}
G_{\rm peak}=\frac{e^2}{\pi\hbar}
\frac{4\Gamma_L\Gamma_R}{(\Gamma_L+\Gamma_R)^2}
\frac{3\pi^2}{8}\left[\ln\frac{\hbar}{\tau T_K}\right]^{-2}.
\label{GRG}
\end{equation}
The width of the conductance peak is $V^*\sim\hbar/e\tau$.
Here the Kondo temperature $T_K$ is defined as \cite{Haldane}
\begin{equation}
T_K=\gamma\sqrt{\frac{(\Gamma_L+\Gamma_R) U}{\pi}}
\exp\left[-\frac{\pi \tilde{E}_d}{2(\Gamma_L+\Gamma_R)}\right]\;,
\label{TK}
\end{equation}
with $\gamma\sim 1$.

At $\hbar/\tau\gg T_K$, one can expand Eq.~(\ref{GRG}) into the series
of powers of $[(\Gamma_L+\Gamma_R)/\tilde{E}_d]\ln(D_0\tau/\hbar)$.
The zero-order term of the series is the conductance calculated in the
Born approximation, and the next term yields the lowest order Kondo
correction given by Eq.~(\ref{Gt}). At $\hbar/\tau\lesssim T_K$, we
expect, in the spirit of the renormalizability of the Kondo problem,
that the $(\ln x)^{-2}$ function in Eq.~(\ref{GRG}) can be replaced by
some universal function ${\cal F}(x)$. In the limit of no irradiation,
${\cal F}(0)=8/3\pi^2$ (unitary limit of the Kondo scattering).

As the frequency $\omega$ of the AC field grows, the rate of the 
decoherence processes increases, and the  height of the zero bias 
conductance peak $G_{\rm peak}$ drops. The  dependence of $G_{\rm 
peak}$ on $\omega$ can be found from Eq.~(\ref{GRG}). For a relatively 
weak AC field, $eV_\omega\ll\tilde{E}_d$, the decoherence time $\tau$ 
is given by Eq.~(\ref{Gammad}) for the frequencies below the 
ionization threshold $\varepsilon_i$, and by 
$\hbar/\tau=(\Gamma/2)(eV_\omega/\tilde{E}_d)^2$ above the threshold.  
One can easily check that the crossover between these two regimes 
leads only to parametrically small relative variations in the peak 
conductance, as $\omega$ increases, say, from $\varepsilon_i/2$ to 
$2\varepsilon_i$.

Another effect of external irradiation on the differential conductance 
$G(V)$ is in producing satellite peaks at 
$eV=n\hbar\omega$. 
If an  external AC field is applied, then, at $eV=n\hbar\omega$, a 
tunneling  electron can hop from a state at the Fermi level in one 
lead to a state at the Fermi level in the other lead, emitting or 
absorbing $n$ photons. Thus at finite bias the external irradiation 
can effectively put a tunneling  electron into zero-bias conditions, 
and  the Kondo anomaly in the  conductance is revived. The height of 
these peaks can be calculated from the  formula (\ref{G}) similarly to 
Eq.~(\ref{Gt}). Here we give the  results for the first satellite 
peak. At low enough irradiation level, $eV_\omega<\hbar\omega$, it is 
sufficient to consider only one-photon processes, accounted for by the 
Hamiltonian (\ref{Ht1}). The resulting correction to the conductance 
at $e|V|$ close to $\hbar\omega$ has the form
\begin{eqnarray}
&&G^{(3)}_{\rm sat}(V)=
3\pi^2 \frac{e^2}{\pi\hbar} \nu^3\left[{\cal J}_{LR}^{\rm
(0)}\right]^2\left[{\cal J}_{RR}^{\rm (0)}+{\cal J}_{LL}^{\rm
(0)}\right]
\left[\frac{eV_\omega}{\tilde{E}_d}\right]^2 \nonumber\\&&\times
\int^0_{-\infty} \!\! dt  \,(-t) \exp (-|t|/\tau_{\rm sat})
\nonumber\\ &&\times
\left(\frac{\pi T}{\hbar}\right)^2
\frac{\cos [(eV/\hbar+\omega)t]
+\cos [(eV/\hbar-\omega)t]}
{\sinh^2(\pi Tt/\hbar)+(T/D_0)^2}\;.
\label{Gtsat}
\end{eqnarray}
When $eV\ne\pm\,\hbar\omega$, the cosine functions cut off the 
logarithmic  divergences. However, when $eV\to\pm\,\hbar\omega$, one 
of the two  cosine terms becomes essentially constant [cf. 
Eq.~(\ref{Gt}) at $V\to  0$], and the differential conductance has a 
peak again. At $T\to 0$,  the height of the conductance peak is 
determined by the spin  decoherence rate $\hbar/\tau_{\rm sat}$. We 
must mention that $\tau_{\rm sat}$ may be significantly shorter than 
$\tau$ given by Eq.~(\ref{Gammad}). The time $\tau$ characterizes the 
spin decoherence at zero bias, whereas the satellite corresponds to a 
finite bias $eV=\pm\hbar\omega$. In this case, the spin decoherence 
occurs mostly due to the tunneling of electrons through the dot (see 
Fig.~\ref{fig:trans}B, and also \cite{MeirEtal93}). The rate of this 
process is given by
\begin{equation}
\frac{\hbar}{\tau_{\rm sat}}=\frac{1}{2\pi}\,\hbar\omega\,
\frac{\Gamma_L\Gamma_R}{\tilde{E}_d^2}\;.
\label{Gammasat}
\end{equation}
Equations (\ref{Gtsat})--(\ref{Gammasat}) yield the formula for the
satellite peak shape, provided $\hbar/\tau_{\rm sat}\gg T_K$.
The shape of the satellite peak in the conductance is given by
\begin{eqnarray}
&&G_{\rm sat}(eV - \hbar\omega)=
\frac{3}{\pi}\frac{e^2}{\pi\hbar}
\frac{\Gamma_L\Gamma_R}{(\Gamma_L+\Gamma_R)^2}
\left[\frac{eV_\omega}{\tilde{E}_d}\right]^2
\nonumber\\
&&\times\left[\frac{\Gamma_L+\Gamma_R}{\tilde{E}_d}\right]^3
\ln\frac{D_0}
{\sqrt{(\hbar/\tau_{\rm sat})^2+(eV - \hbar\omega)^2}}\;,
\label{Gsat}
\end{eqnarray}
and its width  is of the order of $\hbar/e\tau_{\rm
sat}$.

At $T\to 0$, {\em i.e.}, when the unitary limit of tunneling is 
approached, the formation of the satellite peaks is best viewed as 
redistribution of the Kondo anomaly between the elastic tunneling 
processes and the tunneling with absorption/emission of 
photons\cite{TienGordon}. This transfer of spectral weight reduces the 
height of the zero-bias conductance  peak\cite{LopezEtal98}. To 
compare this mechanism with the spin-flip  cotunneling, we note that 
the redistribution of the Kondo anomaly  results from the changes in 
the {\em single-particle} dynamics. To produce a significant deviation 
of the zero-bias conductance from the unitary limit in this way, one 
therefore needs to apply an AC field with amplitude 
\begin{displaymath}
V_\omega\sim\frac{\tilde{E}_d}{e}\;.
\end{displaymath}

The spin-flip cotunneling directly affects the many-body state which 
produces the Kondo anomaly. Due to the fragility of this many-body 
state, it can be destroyed by a relatively weak AC field;  the Kondo 
effect is suppressed already at 
\begin{displaymath}
\frac{\hbar}{\tau(V_\omega)}\gtrsim T_K\;,  
\end{displaymath}
with $\tau(V_\omega)$ given by Eq.~(\ref{Gammad}). 
Comparing  these two conditions on $V_\omega$, we find that 
the decoherence  yields the 
leading effect of AC field on the  zero-bias DC conductance starting 
from parametrically small frequencies, $\hbar\omega>T_K(E_d/\Gamma)^2$ 
of the AC field.

In conclusion, we have demonstrated that the irradiation suppresses 
the DC Kondo conductance across a quantum dot.  This suppression is an 
essentially non-perturbative phenomenon.  Irradiation brings 
decoherence into the spin dynamics of the dot, even  if the photon 
energy is insufficient to ionize the  dot. Finite lifetime of the 
Kondo resonance, resulting from the irradiation-induced decoherence, 
is the main cause of the suppression of the Kondo effect. For 
suppression to occur, it is sufficient that the spin decoherence time 
$\tau$, given by Eq.~(\ref{Gammad}), is shorter than characteristic 
scale $\hbar/T_K$ defined by the Kondo temperature $T_K$ 
[Eq.~(\ref{TK})].  The spin decoherence leads to saturation of the 
low-temperature conductance at $\tau\lesssim\hbar/T$. The condition 
$\tau\lesssim\hbar/T_K$ is readily satisfied at a relatively small 
amplitude of the AC field, when the redistribution of the differential 
conductance from the zero-bias peak to the satellite peaks is 
negligible.

The work at the University of Minnesota was supported by NSF Grant DMR 
97-31756. LG acknowledges the hospitality of the Delft University of 
Technology. LG and AK acknowledge also the hospitality of Institute of 
Theoretical Physics supported by NSF Grant PHY 94-07194 at University 
of California at Santa Barbara, where a part of the work was 
performed. The authors are grateful to L.P.~Kouwenhoven, 
D.~Goldhaber-Gordon and Y.~Meir for useful discussions.

\end{multicols}

\begin{references}

\bibitem{GoldhaberEtal98} D. Goldhaber-Gordon {\em et al.}, Nature
{\bf 391}, 156 (1998).
\bibitem{CronenwettEtal98} S.M. Cronenwett, T.H. Oosterkamp,
L.P. Kouwenhoven, Science {\bf 281}, 540 (1998).
\bibitem{Weis}J. Schmid, J. Weis, K. Eberl, K. von Klitzing,
Physica {\bf B 256--258}, 182 (1998).
\bibitem{HershfieldEtal91} S. Hershfield, J.H. Davies, J.W.
Wilkins, Phys. Rev. Lett. {\bf 67}, 3720 (1991).
\bibitem{MeirEtal93}Y. Meir, N.S. Wingreen, P.A. Lee, Phys. Rev. Lett.
{\bf 70}, 2601 (1993); N.S. Wingreen, Y. Meir, Phys. Rev. B {\bf
49}, 11 040 (1994).
\bibitem{Burstein}J. Appelbaum, Phys. Rev. Lett. {\bf 17}, 91 (1966);
Phys. Rev. {\bf 154}, 633 (1967); L.Y.L. Shen, J.M. Rowell, Solid
State Commun. {\bf 5}, 189 (1967); Phys. Rev. {\bf 165}, 566 (1968).
\bibitem{HettlerSchoeller95} M.H. Hettler, H. Schoeller,
Phys. Rev. Lett. {\bf 74}, 4907 (1995).
\bibitem{Ng96} T.-K. Ng, Phys. Rev. Lett. {\bf 76}, 487
(1996).
\bibitem{NordlanderEtal98} P. Nordlander, N.S. Wingreen, Y.
Meir, D.C. Langreth, cond-mat/9801241.
\bibitem{LopezEtal98} R. L\'{o}pez, R. Aguado, G. Platero,
C. Tejedor, Phys. Rev. Lett. {\bf 81}, 4688 (1998).
\bibitem{GoldinAvishai97} Y. Goldin, Y. Avishai, Phys. Rev. Lett. {\bf
81}, 5394 (1998).
\bibitem{Kouwenhoven} L.P. Kouwenhoven {\em et al.}, Phys. Rev. B {\bf
50}, 2019 (1994); Phys. Rev. Lett. {\bf 73}, 3443 (1994).
\bibitem{SchriefferWolff66} J.R. Schrieffer, P.A. Wolff,
Phys. Rev. {\bf 149}, 491 (1966).
\bibitem{Haldane} F.D.M. Haldane,
J. Phys. C {\bf 11}, 5015 (1978).
\bibitem{Anderson} P.W. Anderson, J. Phys. C {\bf 3}, 2436 (1970).
\bibitem{TienGordon} P.K. Tien, J.P. Gordon, Phys. Rev. {\bf 129}, 647 (1963).
\end{references}
\end{document}